\documentclass{aa}  
\usepackage[varg]{txfonts}
\usepackage{graphicx}
\usepackage{lipsum}
\usepackage{subcaption}         
\usepackage{lscape}             
\usepackage{placeins}           
                                
\begin{document}

   \title{A comparison of pendulum models for large-amplitude longitudinal prominence oscillations}

   \author{I\~nigo Arregui\inst{1,2}}

   \institute{Instituto de Astrof\'{\i}sica de Canarias, E-38205 La Laguna, Tenerife, Spain\\
              \email{iarregui@iac.es}
         \and
            Departamento de Astrof\'{i}sica, Universidad de La Laguna, E-38206 La Laguna, Tenerife, Spain\\ }

   \date{Received September 30, 20XX}

 
  \abstract{
 Large-amplitude prominence oscillations offer diagnostic information relevant to understanding the magnetic and plasma structure of solar prominences. Accurate prominence seismology requires the use of reliable models. The so-called pendulum model for large-amplitude longitudinal prominence oscillations has demonstrated robustness against observations and numerical simulations. Recent improvements  have extended the model to situations with non-uniform gravity, thus leading to corrections that have implications for the inference of the magnetic field strength. In this study we quantify how the different model predictions given by the original and extended pendulum models impact the inference of the minimum magnetic field strength derived from the observed periods of large-amplitude longitudinal prominence oscillations. The analysis we conducted follows a Bayesian approach to solve the inference problem and assess the absolute and relative plausibilities of the two considered models in explaining the observed data, with their uncertainty. We find that the Bayesian solution to the inference problem provides well-constrained posteriors for the minimum magnetic field strength. However, the solutions from each adopted model differ, with differences increasing with the oscillation period. A model comparison analysis results in the extended model being more plausible in the full range of observed periods. However, the magnitude of the Bayes factor is not large enough to determine whether there is positive evidence supporting any of the models. We suggest computing model-averaged posteriors as the most reasonable solution to the inference problem.

 }  
   \keywords{Magnetohydrodynamics (MHD) --
                Methods: statistical --
                Sun: filaments, prominences --
                Sun: oscillations
               }
   \maketitle
\nolinenumbers

\section{Introduction}
Solar prominences are highly dynamic structures, and ground- and space-based observations obtained in the past five decades have shown clear evidence of their oscillatory dynamics \citep{arregui18b}. The oscillations are commonly interpreted in terms of standing or propagating magnetohydrodynamic waves. Based on their velocity amplitude, prominence oscillations are classified into small and large amplitude events \citep{oliver99,oliver02}. Large amplitude oscillations involve a significant part of the prominence structure and display velocity amplitudes greater than 20 km s$^{-1}$. They are excited by episodic energetic disturbances, such as Moreton or EIT waves \citep{eto02,okamoto04},  shock waves \citep{shen14}, nearby jets, and subflares or flares \citep{jing03,li12}. A particular class of large-amplitude prominence oscillations are those in which the prominence material undergoes periodic motions along the longitudinal axis of the structure \citep{jing03,jing06,vrsnak07,zhang12,luna14, zhang17,luna18}. These so-called large-amplitude longitudinal oscillations (LALOs) have periods between 50 and 160 min, damping times of 120 to 600 min, and velocity amplitudes in the range of 30\,--\,100 km s$^{-1}$. 

A number of theoretical models have been proposed in order to explain the observed properties of LALOs in prominences \citep[see e.g.][]{kleczek69,vrsnak07,jing06,luna12}. Among them, the so-called pendulum model by \cite{luna12} has been successful in explaining the oscillations in terms of a restoring force due to the projected gravity in the tube where the threads oscillate and their damping as being a consequence of the steady accretion of mass onto the threads by thermal non-equilibrium processes. The validity of the pendulum model was tested by \cite{zhang12} using Hinode observations and numerical simulations and by \cite{luna16}, who performed 2D non-linear time-dependent simulations of large-amplitude longitudinal oscillations in a dipped magnetic structure and found good agreement between the numerical results and the \cite{luna12} theoretical model. Model predictions were observationally tested for events in the GONG catalogue of filament oscillations by \cite{luna18}. The robustness of the model provides a diagnostic tool to perform prominence seismology in order to infer the curvature radius of the magnetic field lines and the magnetic field strength.

The \cite{luna12} pendulum model makes two main simplifying assumptions: the consideration of uniform gravity and the semi-circular geometry of the supporting flux tubes. \cite{luna22} relaxed these assumptions and extended the model to situations with non-uniform gravity and different flux-tube geometries, with semi-circular, semi-elliptical, and sinusoidal dips. \cite{luna22} find that the spatial variation of the solar gravity introduces a correction in the equations governing longitudinal oscillations and modifies the pendulum-model approximation.  The correction becomes significant for oscillation periods above 60 min, which has implications for the inference of the magnetic field strength. The gravity correction has the interesting effect of introducing a cut-off period such that longitudinal oscillations must have a period below 167 min. On the other hand, the pendulum model turns out to be quite robust and valid for non-circular dips, and the corrected pendulum model provides a good estimate of the radius of curvature at the bottom of the dips for any flux-tube geometry.

In this study, we quantify how the differences in the predictions  by the original \citep{luna12} and the extended \citep{luna22} pendulum models impact the corresponding inferences of the minimum magnetic field strength in prominences.  We adopted a Bayesian approach to solve the inference problems and to quantify the level of evidence in favour of each of these models in view of the observed data and their uncertainty. 

\section{Analytical pendulum models}
In the original pendulum model by \cite{luna12}, the gravity projected along the magnetic field provides the restoring force of longitudinal oscillations. The oscillation period, $P_0$, is given by

\begin{equation}\label{pendulum}
P_0 = 2\pi\sqrt{\frac{R}{g_0}},
\end{equation}
with $R$ as the radius of curvature of the dipped portion of the field lines and $g_0= 274$ m s$^{-2}$ as the solar gravitational acceleration, which is assumed to be uniform.

Because the magnetic structure is self-supporting, the magnetic tension of the dipped part of the tubes must be larger than the weight of the threads, leading to the condition

\begin{equation}\label{condition1}
\frac{B^2}{\mu_0 R} \geq m n g_0 ,
\end{equation}
where $B$ is the magnetic field strength at the bottom of the dip, $\mu_0$ is the magnetic permeability of free space, $n$ is the particle number density, and $m=1.27\, m_{\rm p}$ is the mean particle mass, with  m$_{\rm p}$ as the proton mass. The combination of Equation~(\ref{pendulum}) with the condition~(\ref{condition1}) leads to a constraint for the minimum magnetic field strength at the bottom of the dip \citep{luna12}:

\begin{equation}\label{b_orig}
B\, \mbox{[G]}\geq 26 \left(\frac{n}{10^{11} \mathrm{cm}^{-3}}\right)^{1/2} P_0 \,\mbox{[hr]}.
\end{equation}
Considering the condition of equality in Equation~(\ref{condition1}) enables us to formulate our model $M_1$ for the period of  longitudinal oscillations as a function of the two-parameter vector 
$\mbox{\boldmath$\theta$}=\{B,n\}$ proposed to explain observed data $D=\{P\}$ as

\begin{equation}\label{pm1}
P _{M_1} (n, B) = \frac{B}{f(n)},  \,\, \mbox{with} \,\, f(n)= 26\left(\frac{n}{10^{11} \mathrm{cm}^{-3}}\right)^{1/2}.
\end{equation}
For magnetic field strength values in the range $B\in[1,100]$ G and particle density values in the range $n\in[10^9,10^{11}]$ cm$^{-3}$,  $f(n)\in[2.6,26]$, the model $M_1$ predicts periods in the range $P\in[2.3, 2307]$ min. 

The robustness of this simple model has been assessed against numerical non-linear time-dependent simulations by \cite{luna16}, who showed that for even relatively weak magnetic fields, the back reaction of the magnetic structure to the mass motions does not significantly affect the predictions of the simple pendulum model. Nevertheless, model $M_1$ has a number of shortcomings, such as the assumption of uniform gravity and semi-circular dips.  \cite{luna22} find that for oscillation periods above $\sim$50 min, there are some differences between the predictions of the simple pendulum model and those from models that consider non-uniform gravity and non-circular dips.

\begin{figure*}[t]
\begin{center}
	\includegraphics[scale=0.5]{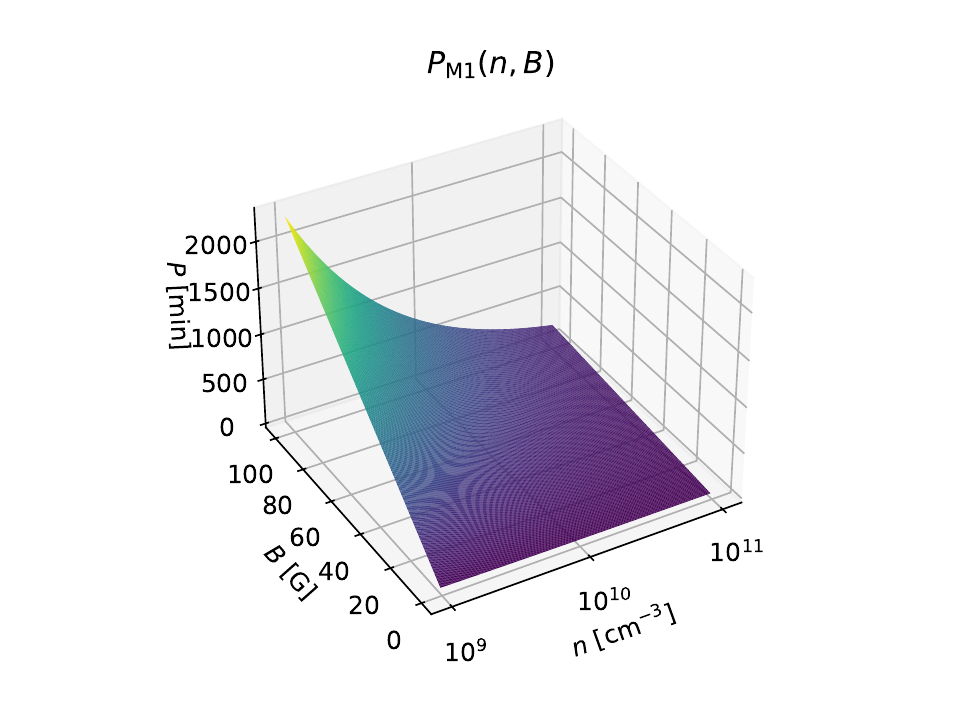}
	 \includegraphics[scale=0.5]{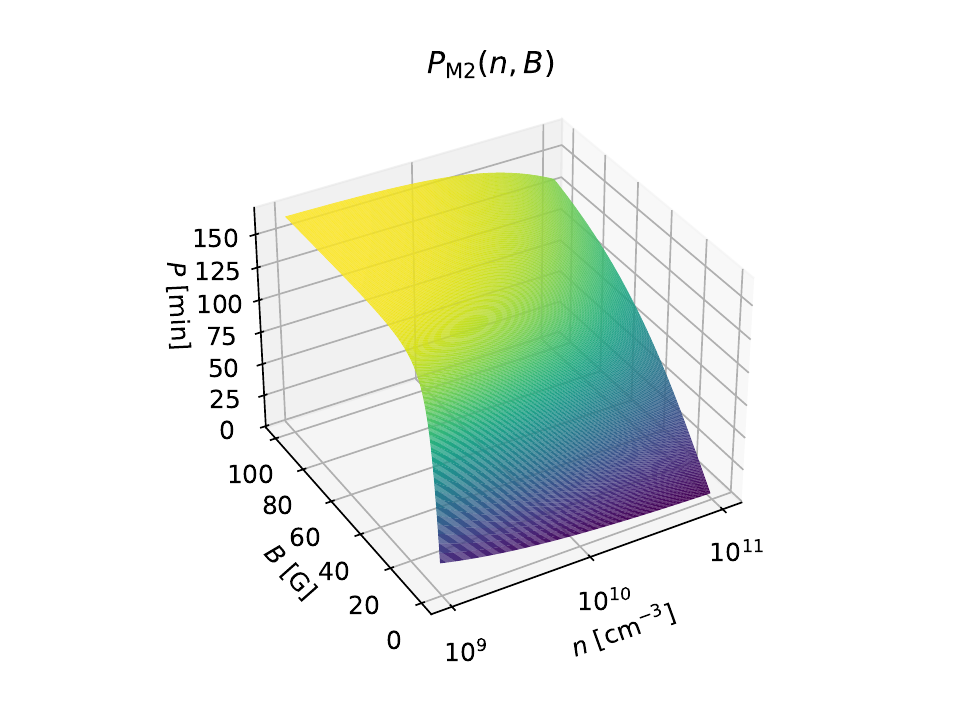}\\
	 \includegraphics[scale=0.4]{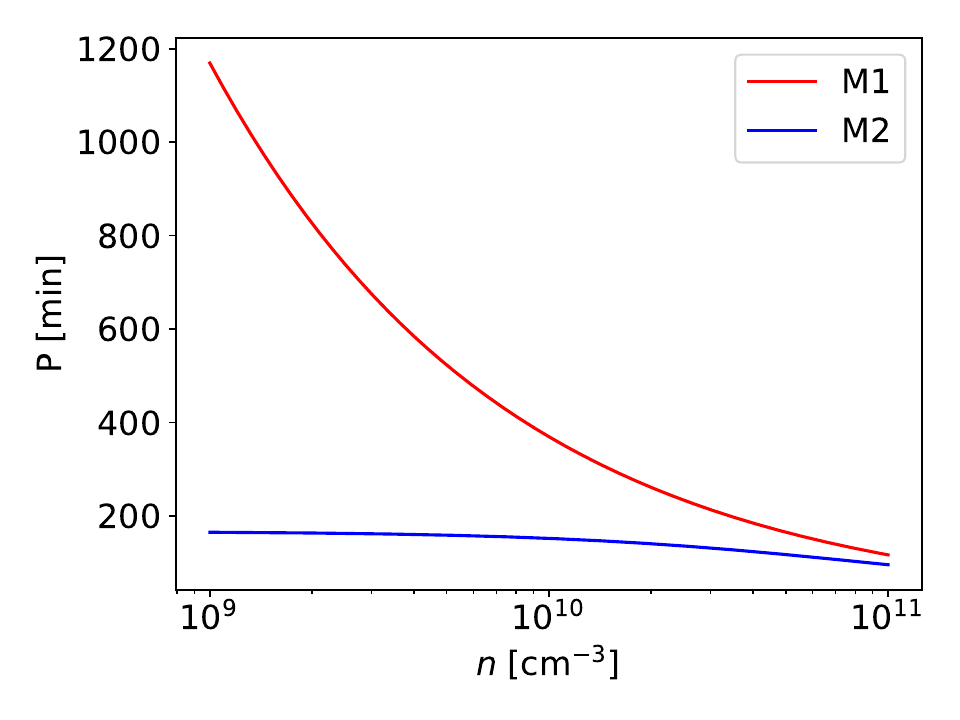}
	 \includegraphics[scale=0.4]{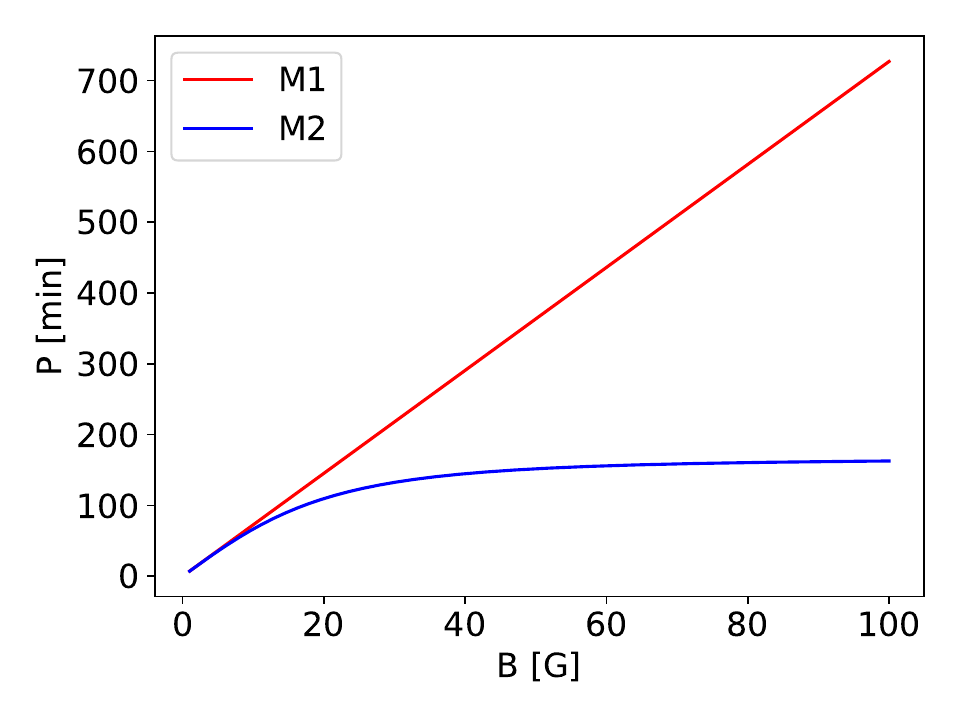}\\
\caption{Top: Surface plots for the period of longitudinal oscillations under model $M_1$ (left, given by Equation~[\ref{pm1}]) and model $M_2$ (right, given by Equation~[\ref{pm2}]) as a function of particle density and magnetic field strength. Bottom: Cuts along a given value of magnetic field strength (left, $B=50$ G) and particle density (right, $n= 10^{10}$ cm$^{-3}$). The solutions have been computed on a uniform 2D grid with $N_{n} = N_{B} = 800$ points. }
\label{fig:models}
\end{center}
\end{figure*}

For non-uniform gravity, \cite{luna22} note that the projection of gravity along the field lines changes due to changes in the magnetic field direction and the intrinsic variation of the direction of gravity. Assuming a semi-circular geometry, \cite{luna22} obtained a formally equivalent expression to Equation~(\ref{pendulum}) in the form

\begin{equation}\label{pendulum1}
P = 2\pi\sqrt{\frac{R_{\rm eq}}{g_0}}, \mbox{\hspace{0.5cm}} \mbox{with}\mbox{\hspace{0.5cm}} \frac{1}{R_{\rm eq}} = \frac{1}{R} + \frac{1}{R_\odot}
\end{equation}
in terms of an equivalent radius of curvature ($R_{\rm eq}$) defined by the harmonic sum of  the radius of curvature of the flux tube and the curvature of the solar surface ($R_\odot$).
According to this expression, the solar-surface curvature introduces a small correction to Equation~(\ref{pendulum}) for periods of around one hour, but there is a significant deviation for longer periods \citep{luna22}.

An interesting consequence of the extended model is the existence of a maximum pendulum period corresponding to the limit of $R\rightarrow \infty$ in which there is no dip, and hence, no support against gravity is possible. This cut-off period is given by

\begin{equation}\label{pcutoff}
P_\odot = 2\pi\sqrt{\frac{R_\odot}{g_0}} = 167 \,\mbox{min}.
\end{equation}

Equation~(\ref{pendulum1}) yields a new relation between the curvature radius and the period and thus a new condition for the support of the prominence material:

\begin{equation}\label{condition2}
B \geq \frac{B_{\rm old}}{\sqrt{1-\left(\frac{P}{P_{\odot}}\right)^2}},
\end{equation}
with $B_{\rm old}$ given by Equation~(\ref{b_orig}).

The condition of equality in Equation~(\ref{condition2}) enabled us to formulate our model $M_2$ for the period of longitudinal oscillations as a function of the same two-parameter vector as before:

\begin{equation}\label{pm2}
P_{M_2}(n,B)  = P_{\odot} \frac{B }{\sqrt{B^2 + P_{\odot}^2 f(n)^2}}.  
\end{equation}
Considering the same ranges in the magnetic field strength and particle density as above, model $M_2$ predicts periods between $P\in[2.3,166]$ min, with an upper limit very close to the cut-off period. 

Model predictions for the period of longitudinal oscillations as a function of plasma density and magnetic field strength in given ranges under models $M_1$ and $M_2$ are shown in Figure~\ref{fig:models}. The surface plots in Figures~\ref{fig:models}  demonstrate the differences in the general structure of the predictions between the models. Model $M_1$ appears to predict unlimited periods as the plasma density decreases and the  magnetic field strength increases (see Figure~\ref{fig:models}, top-left). On the contrary, the cut-off on the period imposed by model $M_2$ and given by Equation~(\ref{pcutoff}) can be appreciated in Figure~\ref{fig:models}, top right. Cuts along particular values of magnetic field strength and density in the bottom panels of Figure~\ref{fig:models} show quantitative differences in model predictions that become more important with decreasing density and increasing magnetic field strength (increasing oscillation periods).

\section{Bayesian methodology}\label{sec:method}
In order to quantify the impact of the model prediction differences on the inference of the minimum magnetic field strength, we adopted a Bayesian framework for parameter inference and model comparison. The analysis is based on the use of Bayes's theorem for the inference of a set of parameters, $\mbox{\boldmath$\theta$}$, pertaining, to a model $M$, conditional on observed  data, $D$, as

\begin{equation}\label{bayes}
p(\mbox{\boldmath$\theta$}|D,M) = \frac{p(D|\mbox{\boldmath$\theta$}, M)p(\mbox{\boldmath$\theta$}|M)}{\int p(D|\mbox{\boldmath$\theta$}, M)p(\mbox{\boldmath$\theta$}|M) \, \mathrm{d}\mbox{\boldmath$\theta$}}.   
\end{equation}
In this expression  $p(\mbox{\boldmath$\theta$}|D,M)$ is the posterior probability distribution of the parameters, $p(D|\mbox{\boldmath$\theta$}, M)$ is the likelihood function, and $p(\mbox{\boldmath$\theta$}|M)$ is the prior distribution of the parameters. The denominator is the model evidence, which acts as a normalisation constant in parameter inference and as a relational measure of evidence in model comparison.

\begin{figure*}[t]
\sidecaption
	\includegraphics[width=10cm]{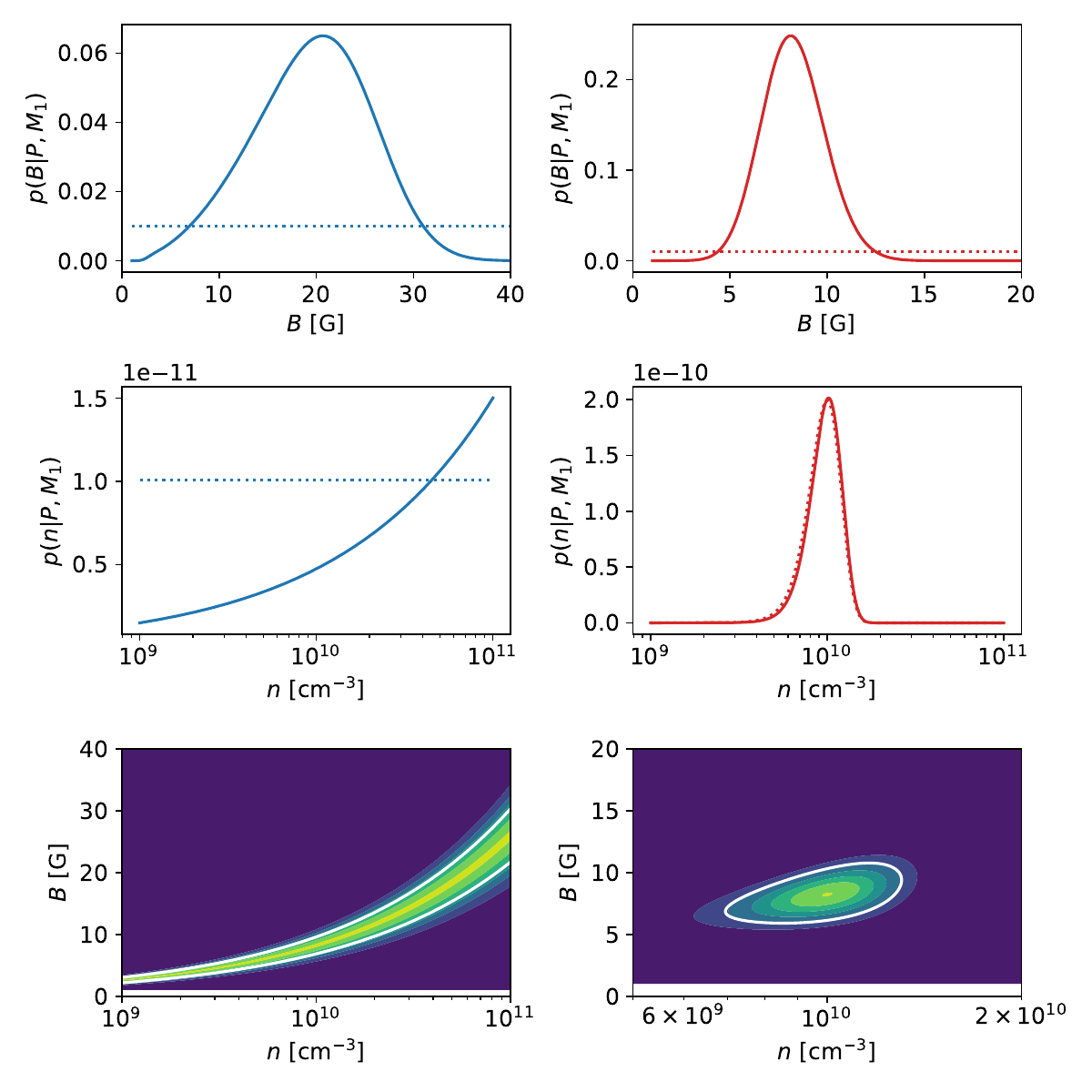}		
\caption{Marginal and joint posterior distributions for the minimum magnetic field strength $B$ and the plasma particle density $n$ for a LALO with period $P=60\pm10$ min under model $M_1$ given by Equation~(\ref{pm1}). The left panels are the results obtained with uniform priors, $\mathcal{U}(B \,\mathrm{[G]}; 1,100)$ and $\mathcal{U}(n$ [cm$^{-3}$]; 10$^9$,10$^{11}$) (blue-dotted lines). The right panels are the results obtained with $\mathcal{U}(B \,\mathrm{[G]}; 1,100)$ and a Gaussian prior on particle density, $\mathcal{G}(n$ [cm$^{-3}$]; $\mu_{n},\sigma_{n})$ (red dotted lines), with $\mu_n=10^{10}$ cm$^{-3}$ and $\sigma_n=0.2\mu_n$. The top and middle panels show the marginal posteriors, and the bottom panels show the joint posteriors, with the white line enclosing the 68$\%$ credible interval. For the magnetic field strength, the median and upper bounds at the 68\% credible intervals are $B=19^{+5}_{-6}$ G for the uniform priors and $B=8^{+1}_{-1}$ G for the Gaussian prior on density. Solutions computed over a 2D grid with $N_{n} = N_{B} = 800$ points.}
\label{fig:inference-example}
\end{figure*}

The prior $p(\mbox{\boldmath$\theta$}|M)$ encodes the information about the parameter vector we are willing to assume before having considered the observed data. In our study, we adopted two types of priors. When a given parameter, $\theta_i$, is assumed to lie in a plausible range with equal probability for all values within the interval, a uniform prior is defined in terms of the lower ($\theta_{i_{\rm min}}$) and upper ($\theta_{i_{\rm max}}$) bounds in the form

\begin{equation}\label{uniformprior}
    \mathcal{U}(\theta_i; \theta_{i_{\rm min}},\theta_{i_{\rm max}})=\begin{cases}
    (\theta_{i_{\mathrm{max}}} - \theta_{i_{\mathrm{min}}})^{-1} & \text{$\theta_{i_{\mathrm{min}}} \le \theta_i\le \theta_{i_{\mathrm{max}}}$} \\ 
    0 & \text{otherwise}.
    \end{cases}
\end{equation}
On the other hand, when some information is available from observations or other sources, a Gaussian density can be used as a prior, with a mean ($\mu_{\theta_i}$) determined by the observed estimate and a standard deviation ($\sigma_{\theta_i}$) given by the uncertainty in the estimate such that

\begin{equation}\label{gaussianprior}
    \mathcal{G}(\theta_i; \mu_{\theta_i},\sigma_{\theta_i}) = \frac{1}{\sqrt{2\pi}\sigma_{\theta_i}}  \exp{\left[\frac{-(\theta_i - \mu_{\theta_i})^2}{2\sigma_{\theta_i}^2}\right]}.
\end{equation}

The function $p(D|\mbox{\boldmath$\theta$}, M)$ in Equation~(\ref{bayes}) measures the distance between a given observation ($D$) and the prediction of the model [$D_M(\mbox{\boldmath$\theta$})$] and assigns a likelihood to the different parameter combinations based on the magnitude of that distance in relation to the uncertainty on the data ($\sigma_{D}$). We assumed a Gaussian likelihood of the form

\begin{equation}\label{likelihood}
    p(D|\mbox{\boldmath$\theta$}, M) = \frac{1}{\sqrt{2\pi}\sigma_{D}}  \exp{\left[\frac{-(D - D_{M}(\mbox{\boldmath$\theta$}))^2}{2\sigma_{D}^2}\right]}.
\end{equation}

The combination of likelihood,  $p(D|\mbox{\boldmath$\theta$}, M)$, and prior information, $p(\mbox{\boldmath$\theta$}|M)$, leads to the full posterior $p(\mbox{\boldmath$\theta$}|D,M)$.  To know how the data constrain one particular parameter, $\theta_i$, one can compute the so-called marginal posterior as

\begin{equation}\label{marginal}
    p(\theta_i|D, M) = \int p(\mbox{\boldmath$\theta$}|D,M) \, \mathrm{d}\theta_1\,...\,\mathrm{d}\theta_{i-1},\mathrm{d}\theta_{i+1}\,...\,\mathrm{d}\theta_N.   
\end{equation}
This integral encodes all the information for the model parameter $\theta_i$ available in the priors and the data. An additional advantage is that it correctly propagates uncertainties in the rest of the parameters to the one of interest.

To compare the evidence in favour of the two models $M_1$ and $M_2$ in view of the observed data, $D$, we employed two measures: the marginal likelihood and the Bayes factor. The marginal likelihood (or evidence) appears in the denominator of Bayes's theorem (Equation~[\ref{bayes}]) and is the integral of the joint distribution over the full parameter space:

\begin{equation}\label{eq:ml}
p(D|M) = \int_{\mbox{\scriptsize\boldmath$\theta$}} p(\mbox{\boldmath$\theta$}, D| M) \, \mathrm{d}\mbox{\boldmath$\theta$} = \int_{\mbox{\scriptsize\boldmath$\theta$}} p(D|\mbox{\boldmath$\theta$},M) \, p(\mbox{\boldmath$\theta$}|M) \, \mathrm{d}\mbox{\boldmath$\theta$}.
\end{equation}
The marginal likelihood quantifies the evidence of a model in relation to its predictive accuracy for observed data. To assess the relative evidence between alternative models in explaining the same observed data, we considered posterior ratios, $p(M_{1}|D)/p(M_{2}|D)$. If the two models are equally probable a priori, $p(M_1) = p(M_2)$, and through application of Bayes's rule, the posterior ratio reduces to the ratio of marginal likelihoods of the two models 

\begin{equation}\label{eq:bf}
BF_{12} = 2\log \frac{p(D|M_1)}{p(D|M_2)} = -BF_{21},
\end{equation}
where the logarithmic scale is used for convenience in the evidence assessment. The Bayes factors defined in Equation~(\ref{eq:bf}) quantify the relative plausibility of each of the two models to explain the same data. The quantitative assessment is made in terms of levels of evidence with the use of an empirical table, such as the one  by \cite{kass95}. For instance, the evidence in favour of model $M_1$ in front of model $M_2$ is inconclusive for values of $BF_{12}$ from zero to two, positive for values from two to six, strong for values from six to ten, and very strong for values above ten. A similar tabulation applies to $BF_{21}$. 

\section{Results}

\subsection{Parameter inference}
We applied the methodology described in Section~\ref{sec:method} to perform the inference of the relevant physical parameters using the period of longitudinal prominence oscillations under the original $M_1$ and extended $M_2$ pendulum models. In both models, the unknown parameters form a 2D parameter vector $\mbox{\boldmath$\theta$}=\{n,B\}$, and the observed data are the period of longitudinal oscillations $D =\{P\}$, with uncertainty $\sigma_D=\Delta P$.

We first considered the original pendulum model, $M_1$, with predictions for the period given by Equation~(\ref{pm1}). To construct the priors, we considered that the magnetic field strength and the particle density vary over given ranges. Two types of prior combinations were used. On one hand, we adopted uniform priors for both parameters within particular ranges, following the definition in Equation~(\ref{uniformprior}). On the other hand, we computed solutions that adopt a uniform prior for $B$ and a Gaussian prior on particle density, following the definition in Equation~(\ref{gaussianprior}). The idea is that if some estimate, $\mu_n$, with error $\sigma_n$ can be obtained from observations, they can be incorporated into the inference process.

\begin{figure}[t]
\begin{center}
         \includegraphics[scale=0.45]{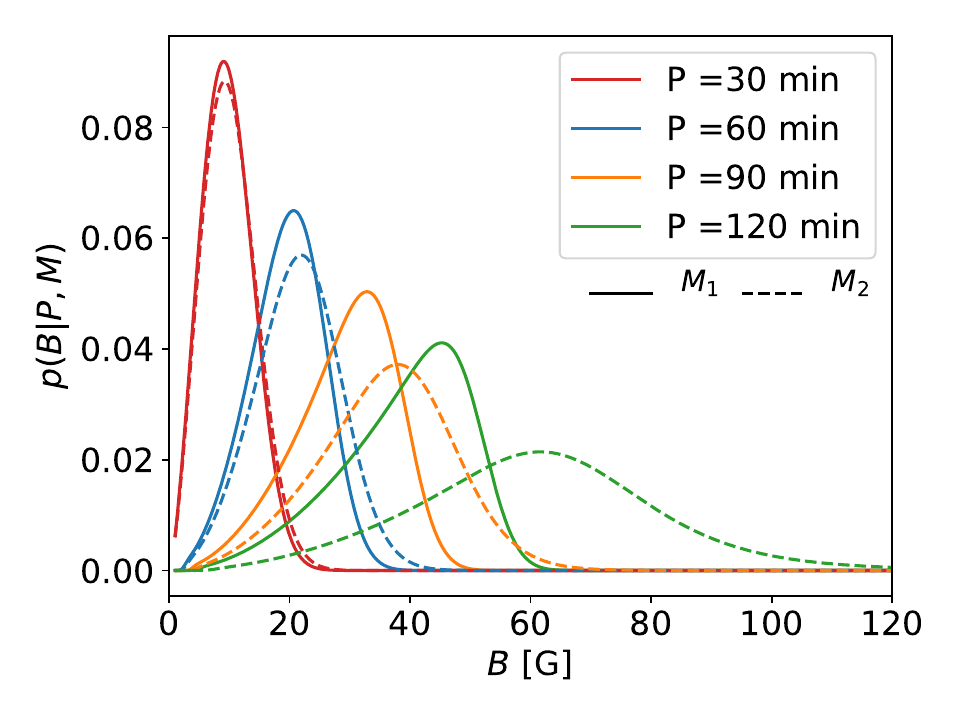}\\
         \includegraphics[scale=0.45]{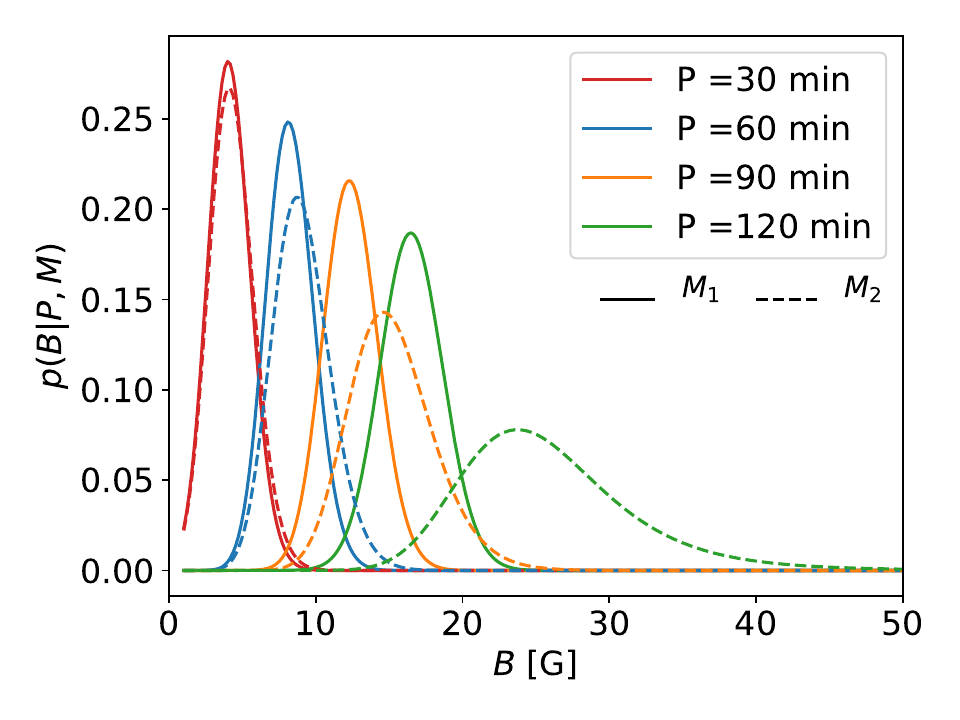}        
\caption{Posterior densities for the minimum magnetic field strength for four values of the period under the original $M_1$ (solid) and the extended $M_2$ (dashed) pendulum models. The top panel displays results obtained with uniform priors, $\mathcal{U}(B \,\mathrm{[G]}; 1,100)$ and $\mathcal{U}(n$ [cm$^{-3}$]; 10$^9$,10$^{11}$). The bottom panel shows the results obtained with $\mathcal{U}(B \,\mathrm{[G]}; 1,100)$ and a Gaussian prior on particle density, $\mathcal{G}(n$ [cm$^{-3}$]; $\mu_{n},\sigma_{n})$, with $\mu_n=10^{10}$ cm$^{-3}$ and $\sigma_n=0.2\mu_n$. The numerical summaries of the posteriors are given in Table~{\ref{table:stats}}. A value of  $\sigma_P = 10$ min and a 2D grid with $N_{n} = N_{B} = 800$ points were considered.}
\label{fig:inferences}
\end{center}
\end{figure}

An example inference result using both prior types and the likelihood function in Equation~(\ref{likelihood}) for a typical value of the period of longitudinal prominence oscillations is shown in Figure~\ref{fig:inference-example}. The results indicate that a well-constrained marginal posterior for the minimum magnetic field strength can be obtained (see top panels of Figure~\ref{fig:inference-example}), even when information about the plasma particle density is limited to being on a given range (top-left panel). The use of uniform priors leads to a marginal posterior for the magnetic field strength distributed within a larger range of values in comparison to the case in which a Gaussian prior on density is adopted.  The plasma particle density cannot be well constrained unless the more informative Gaussian prior is used (see middle panels of Figure~\ref{fig:inference-example}). In that case, the prior and posterior for density are very similar, meaning that seismology is of little help in the inference of this parameter. A situation akin to this was also found in the inference of the magnetic field strength and the plasma density using transverse kink oscillations in coronal loops by \cite{arregui19b} and in prominence threads by \cite{montesolis19} and was also recently pointed out in the study by \cite{baweja26}. The bottom panels in Figure~\ref{fig:inference-example} show the marked differences in the joint posterior distributions that are obtained when using either a uniform or a Gaussian prior for the plasma density.

\begin{table*}
\caption{Inference results.}             
\label{table:stats}    
\centering                        
\begin{tabular}{cc| c c c| ccc}      
\hline\hline               
&Observations&\multicolumn{3}{|c|}{Uniform priors} & \multicolumn{3}{c}{Gaussian prior on $n$}\\
\hline
&P $\pm$ $\Delta P$ [min] & $B_{M_1} [G]$ & $B_{M_2} [G]$ & $B_{\rm avg}$ [G] &$B_{M_1}$ [G] & $B_{M_2} [G]$ &$B_{\rm avg}$ [G]  \\         
\hline                      
\\
  Synthetic & 30 $\pm$ 10 & 9$^{+4}_{-4}$ & 9$^{+4}_{-4}$ &9$^{+4}_{-4}$&3$^{+1}_{-1}$  & 4$^{+1}_{-1}$ & 4$^{+1}_{-1}$\\ \\
      
  Synthetic  &60 $\pm$ 10 & 19$^{+5}_{-6}$ & 21$^{+6}_{-7}$ &20$^{+6}_{-6}$&8$^{+1}_{-1}$  & 8$^{+2}_{-1}$ & 8$^{+1}_{-1}$\\   \\ 
   
   Synthetic &90 $\pm$ 10 & 30$^{+7}_{-9}$ & 36$^{+10}_{-11}$ &33$^{+10}_{-10}$&12$^{+1}_{-1}$  & 14$^{+3}_{-2}$ &13$^{+3}_{-2}$\\    \\
   
  Synthetic  &120 $\pm$ 10 & 40$^{+8}_{-12}$ & 60$^{+19}_{-19}$ &53$^{+20}_{-17}$&16$^{+2}_{-2}$ & 24$^{+6}_{-4}$&23$^{+6}_{-6}$\\  \\ 
  
 \cite{zhang17}  &99 $\pm$ 10  &  33$^{+7}_{-10}$ & 42$^{+11}_{-13}$ &38$^{+12}_{-12}$&27$^{+4}_{-3}$ & 35$^{+7}_{-6}$&32$^{+8}_{-6}$\\
\\     
\hline\hline               
\end{tabular}
\tablefoot{Posterior summaries for different synthetic periods and one observed event under the two considered models and the two adopted sets of priors. The corresponding marginal posteriors are displayed in Figure~\ref{fig:inferences} for the synthetic observations and in Figure~\ref{fig:average} for the event observed by \cite{zhang17}. Posterior summaries are given as the median and upper and lower bounds at the 68\% credible interval. We note that the Gaussian prior in density for the synthetic cases is $\mathcal{G}(n$ [cm$^{-3}$]; $\mu_{n},\sigma_{n})$, with $\mu_n=10^{10}$ cm$^{-3}$ and $\sigma_n=0.2\mu_n$, while for the observed event, it is $\mathcal{G}(n$ [cm$^{-3}$]; $\mu_{n},\sigma_{n})$, with $\mu_n=4.25\times10^{10}$ cm$^{-3}$ and $\sigma_n=0.2\mu_n$.}
\end{table*}

We subsequently performed the inference for a set of four different synthetic periods within the range of typically observed values \citep{luna18} and compared the outcome with the results obtained with the models $M_1$ and $M_2$.  The results displayed in Figure~\ref{fig:inferences} show that as the period of longitudinal oscillations increases, the posteriors (and hence the estimates) for the minimum magnetic field strength shift towards larger values, in logical agreement with the forward predictions in Equations~({\ref{pm1}}) and (\ref{pm2}) and Figure~\ref{fig:models}.  The posteriors for the case of a Gaussian prior on density (bottom panel) are more symmetric. The posteriors obtained with the extended pendulum model, $M_2$, which considers non-uniform gravity, result 
in higher values of the inferred magnetic field strength in comparison to the posteriors obtained with the original pendulum model, $M_1$. The differences are small for the shortest periods, 30 and 60 min, but they are clearly discernible for the longest periods, 90 and 120 min. This result is independent of the type of prior information being employed. Table~\ref{table:stats} shows posterior summaries corresponding to the inference results for all the considered synthetic periods, models, and prior assumptions. 

\begin{figure}[t]
\begin{center}
         \includegraphics[scale=0.45]{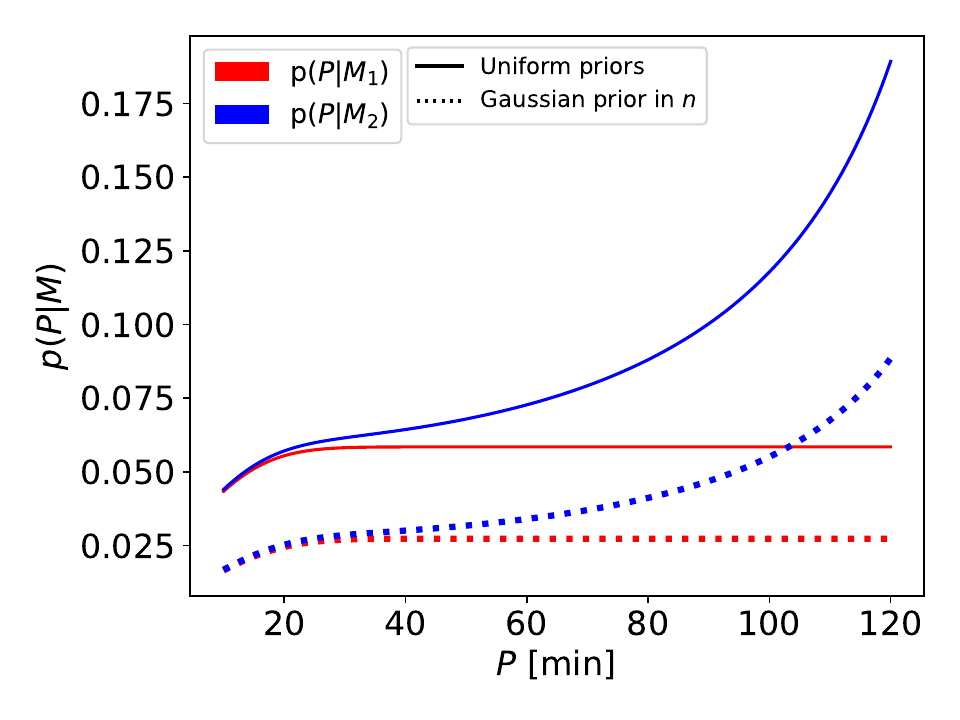}\\
                  \includegraphics[scale=0.45]{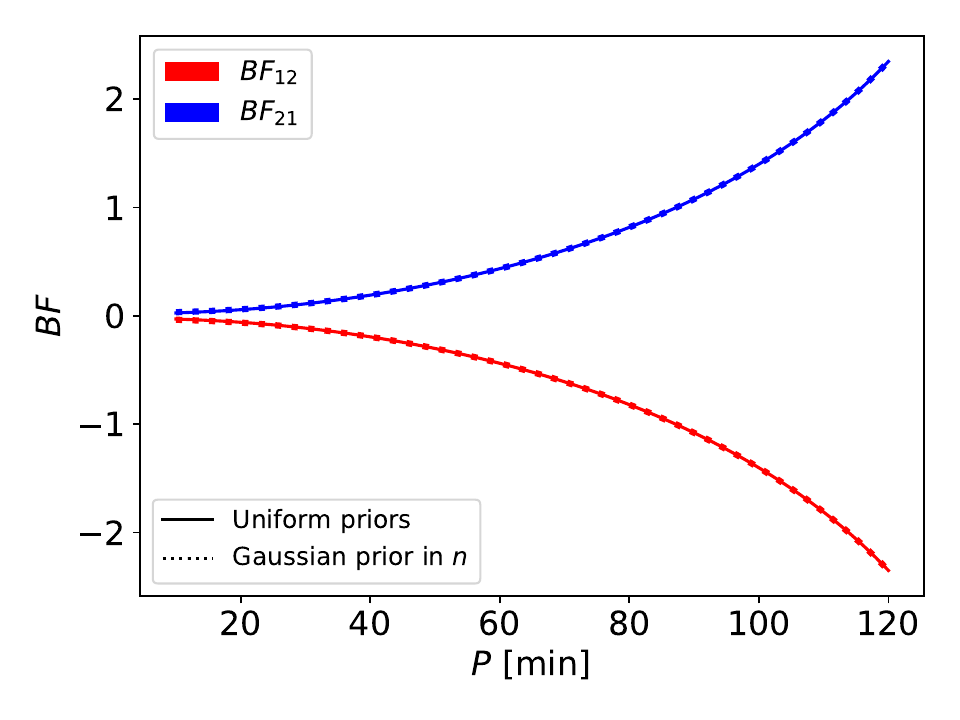}

\caption{Top: Marginal likelihood for models $M_1$ and $M_2$ as a function of the period, with uncertainty, computed using Equation~(\ref{eq:ml}). Bottom: Bayes factors for the relative plausibility between $M_1$ and $M_2$ as a function of the oscillation period, computed using Equation~(\ref{eq:bf}). The solid (dotted) line corresponds to the use of uniform (Gaussian) priors. A value of  $\sigma_P = 10$ min and a 2D grid with $N_{n} = N_{B} = 800$ points were considered.}
\label{fig:ml}
\end{center}
\end{figure}

\subsection{Model comparison}
We have shown that the Bayesian solution to the inference problem enables us to obtain well-constrained posteriors for the magnetic field strength from observed periods of large-amplitude longitudinal prominence oscillations. However, the obtained posteriors differ depending on the particular model adopted: original or extended pendulum model. The differences are small for the shortest period oscillations, but they become more marked as we increase the oscillation period and cannot be neglected for periods above  $\sim$ 50 min. The question then arises about which inference, that from model $M_1$ or model $M_2$, is to be preferred. The Bayesian approach offers two tools to quantify the absolute and relative plausibility of the models in view of data: the marginal likelihood and the Bayes factor.

We first computed the marginal likelihood for the two models under consideration in a range that considers observed periods from 10 to 120 min (see histogram of data in Figure 24 of \citealt{luna18}) by applying Equation~(\ref{eq:ml}) to each model. The integral in this equation takes into account all the possible parameter combinations that could produce the observed data, considering their prior probability, their likelihood, and the uncertainty of the data. Hence, all the available information is used in a consistent way to quantify how good each model is at producing a particular observed period. The top panel in Figure~\ref{fig:ml} shows the obtained results. The magnitude of the marginal likelihood is a measure of the global plausibility of each particular model in explaining those periods. The results indicate that the marginal likelihood for model $M_2$ is larger than the marginal likelihood for model $M_1$ for all periods. For the shortest considered periods, below $\sim$ 25 min, the marginal likelihood for both models is very similar. As the period increases, the marginal likelihood for the extended pendulum model, $M_2$, is clearly larger than the marginal likelihood for the original pendulum model, $M_1$. This is true regardless of the prior assumptions. 

Next, we assessed the relative plausibility of model $M_1$ against model $M_2$ (and vice versa) by computing the Bayes factor $BF_{12}$ ($BF_{21}$) through application of Equation~(\ref{eq:bf}). The bottom panel in Figure~\ref{fig:ml} displays the obtained results. The magnitude of the Bayes factor $BF_{21}$ is positive and increases with increasing period. This means that the extended pendulum model, $M_2$, is more plausible than the original pendulum model, $M_1$, for all periods in the observed period range of LALOs. This is true regardless of the prior assumptions used, and indeed, the Bayes factors are mostly independent of the prior assumptions. We note, however, that the magnitude of the Bayes factors is almost always below two ($BF_{21} \sim 2$ for $P=118$ min), which marks the limit above which positive evidence in favour of $M_2$ against $M_1$ can be established. The condition $BF_{21}> 2$, leading to positive evidence of $M_2$, is thus reached for periods above those observed.

\subsection{Model averaging}
Since the obtained Bayes factors are not large enough to determine whether there is positive evidence in favour of the models under comparison, the question as to which model-based inference ($M_1$ or $M_2$) is preferred remains. The solution to this problem in the Bayesian framework is to consider model averaging. Bayesian model averaging enables us to obtain parameter constraints that account for the uncertainty regarding the models under consideration. In our case, it consists of combining the posteriors obtained for each pendulum model to obtain a model-averaged posterior, weighted with the evidence of each model.

The model-averaged posterior distribution for the minimum magnetic field strength, conditional on the observed period and weighted with the probability of our set of two models ($M_1$ and $M_2$) is given by

\begin{eqnarray}
p_{\rm avg}(B|P) &=& p(B|P,M_1)\, p(M_1|P) + p(B|P,M_2)\, p(M_2|P)\nonumber\\
&=& p(M_1|P) \left[BF_{11} p(B|P, M_1) +  BF_{21} p(B|P, M_2)\right],\nonumber\\
\end{eqnarray}
where in the second equality we adopt the original pendulum model ($M_1$) as the reference model and replace the model's posterior probabilities by the Bayes factors $BF_{11}$ and $BF_{21}$, with respect to the reference model. Obviously, $BF_{11}=1$. The posterior probability for the reference model can be calculated by considering that the sum of the probabilities for the two models must be unity, and thus

\begin{equation}
p(M_1|P) = \frac{1}{1+BF_{21}}.
\end{equation}

\begin{figure}[t]
\begin{center}
         \includegraphics[scale=0.5]{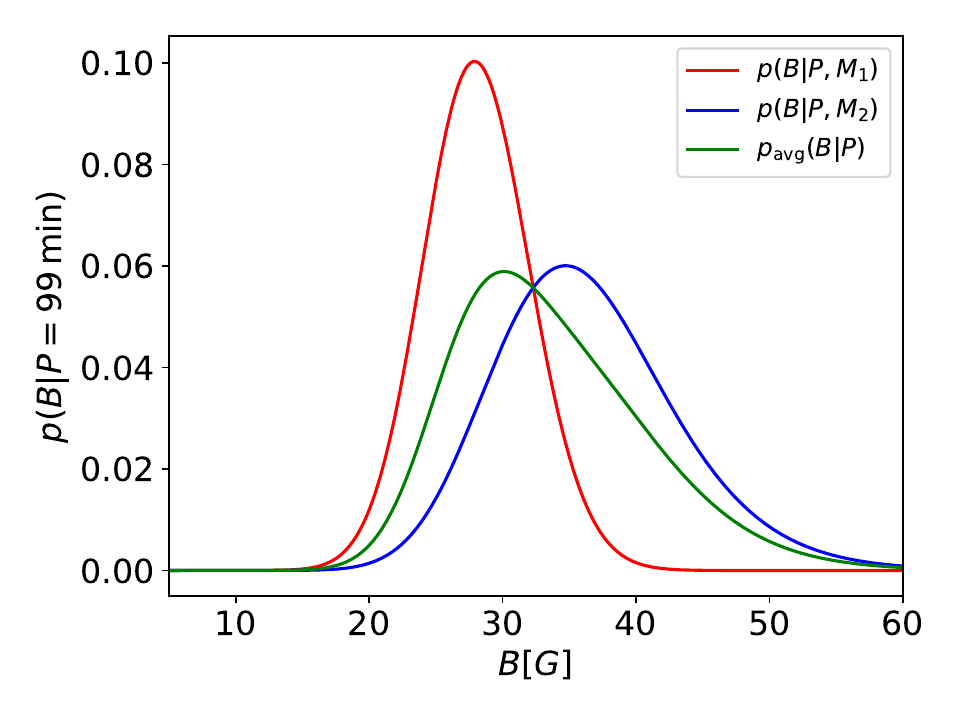}        
\caption{Marginal posteriors under models $M_1$ and $M_2$ and model-averaged posterior for the magnetic field strength for the oscillation with $P= 99$ min analysed by \cite{zhang17}. A uniform prior, $\mathcal{U}(B \,\mathrm{[G]}; 1,100)$, and a Gaussian density prior, $\mathcal{G}(n$ [cm$^{-3}$]; $\mu_{n},\sigma_{n})$, with $\mu_n=4.25\times10^{10}$ cm$^{-3}$ and $\sigma_n=0.2\mu_n$ have been used. The numerical summaries of the posteriors are given in Table{\ref{table:stats}}. A value of  $\sigma_P = 10$ min and a 2D grid with $N_{n} = N_{B} = 800$ points were considered.}
\label{fig:average}
\end{center}
\end{figure}

As a representative example application of Bayesian model averaging, we considered the LALO event reported by \cite{zhang17}, which has a relatively long period of $\sim 99$ min, yielding an estimate for the minimum magnetic field strength of $\sim 28$ G under the original pendulum model. \cite{luna22} report that when taking into account the correction due to the non-uniform gravity model, the estimated magnetic field value is $\sim 35$ G. Figure~\ref{fig:average} shows the Bayesian inference results under the two models, $M_1$ and $M_2$, and the model-averaged posterior for this observed event and when considering a Gaussian prior for the particle density, based on the value assumed by \cite{zhang17} and \cite{luna22}.  The summary statistics for the inference from each model and for the model-averaged posteriors are listed in Table~\ref{table:stats}. The probabilities for each model are $p(M_1|P)=0.34$ and $p(M_2|P)=0.66$. The Bayes factors are $BF_{12}=-1.36 = - BF_{21}$. Model $M_2$ is more plausible than model $M_1$, but the Bayes factor is below the value required to determine positive evidence. The model-averaged posterior lies between the posteriors obtained with the individual models,  $p(B|P,M_1)$ and $p(B|P,M_2)$. 

In situations such as this, the model-averaged posteriors offer the most general inference result that can be obtained.  They embrace all the
available information, i.e. the prior information, the observed data with their uncertainty, and the evidence of each model in view of data, in a fully consistent way.

\section{Summary and conclusions}

We estimated the impact of the differences between the predictions of two pendulum models on the inference of the minimum magnetic field strength in solar prominences based on the observation of LALOs. The two models differ in their consideration of either uniform or non-uniform gravity. We also quantified the plausibility of the models in view of data and their uncertainty.

The original (uniform gravity) and extended (non-uniform gravity) pendulum models exhibit differences in the predicted periods as a function of plasma density and magnetic field strength. The differences become important in the region of the parameter space with the largest magnetic field strengths and the lowest plasma densities, among those considered. The extended model imposes a maximum period beyond which no support against gravity is possible, while the original model predicts periods that grow unbounded for strong magnetic fields.

The Bayesian solution to the inference problem enables well-constrained posteriors to be obtained for the minimum magnetic field strength, even when information on the plasma density is poorly constrained. Having additional knowledge of this physical parameter in the form of prior information enables further constraints on the inference to be made. The inference of the minimum magnetic field strength with the extended model leads to higher values in comparison to the inference with the original pendulum model. The differences are small for short oscillation periods below 60 min but significant for the longest periods above that value.

The marginal likelihood for the extended model is larger than the marginal likelihood for the original pendulum model in the full range of observed periods of LALOs. This means that for the parameter ranges considered in the study, the extended model better predicts the observations. The analysis of the relative plausibility between the two models leads to Bayes factors that support the extended model, but they are not large enough to be considered positive evidence when the models are considered equally, likely a priori. For the considered uncertainty in the observations, the condition of positive evidence is reached for periods $\ge$ 118 min, which are above those observed.

With these results,  the adoption of model-averaged posteriors turns out to be the most sensible solution to the inference problem. Model-averaged posteriors offer the most general inference result that considers all the available information and their uncertainty in a consistent manner. This conclusion may appear awkward from the perspective of classic modelling since a model incorporating a physical ingredient (non-uniform gravity) present in the atmosphere can be regarded as more plausible. This would be strictly true provided we know the value of the physical parameters with certainty. The fact that we are uncertain about the equilibrium parameters and the observed data makes our judgement about the plausibility of the models of probabilistic nature.

\begin{acknowledgements}
This study was supported by project PID2024-156538NB-I00 funded by MCIN/AEI/10.13039/501100011033 and by ``ERDF A way of making Europe". Nere amatxori, Ixiar Uribeetxebarria Erostarbe.
\end{acknowledgements}

\end{document}